\documentclass[fleqn,twoside]{article}
\usepackage{espcrc2}

\usepackage{graphicx}
\usepackage[figuresright]{rotating}

\newcommand{\AmS}{{\protect\the\textfont2
  A\kern-.1667em\lower.5ex\hbox{M}\kern-.125emS}}

\title{Comparing matrix models and QCD lattice data with chemical potential
}

\author{Gernot Akemann\address{%
Service de Physique Th\'eorique, CEA/DSM/SPhT Saclay,\\
Unit\'e associ\'ee CNRS/SPM/URA 2306,
F-91191 Gif-sur-Yvette Cedex, France}%
        \thanks{Supported by a Heisenberg fellowship of the DFG.}
        and
        Tilo Wettig\address{%
Department of Physics, Yale University, New Haven, CT 06520-8120, USA and\\
  RIKEN-BNL Research Center, Brookhaven National Laboratory, Upton, NY
  11973-5000, USA}%
        \thanks{Supported in part by DOE grant DE-FG02-91ER40608.}}
       
\begin{document}

\begin{abstract}
We present a quantitative analysis of the microscopic Dirac spectrum
which is complex in the presence of a non-vanishing quark chemical
potential.  Data from quenched SU(3) lattice simulations for
different volumes $V$ and small values of the chemical potential $\mu$ are
compared to analytical predictions from matrix models. We
confirm the existence of two distinct limits for weakly and strongly
nonhermitian Dirac operators. Good agreement is found 
in both limits, confirming the 
different scaling of chemical potential and eigenvalues with the volume.
\vspace{1pc}
\end{abstract}

\maketitle

\setcounter{footnote}{0}

\section{Introduction}

Lattice QCD at nonzero density remains a difficult topic. Several methods to 
solve the problems related to the complex phase of the Dirac operator 
determinant have recently been suggested: multiparameter reweighting, 
Taylor expansion, and analytic continuation from 
imaginary $\mu$ \cite{FK02}. So far, all three methods 
work only in the vicinity of the phase transition close to the 
temperature-axis. Here, we instead explore the 
$\mu$-axis at zero temperature and small $\mu$. In this region matrix models 
analytically predict the complex microscopic Dirac spectrum \cite{A02}.  
We test these predictions for pure SU(3) gauge theory \cite{AW03}, 
where $\mu\neq0$ is easily implemented.
Although the quenched approximation is problematic \cite{Step96}, 
our test is an important check, as the matrix model remains valid 
and predictive for $N_f>0$ flavors. For example, the   1-flavor
matrix model has been used to test a new factorization algorithm for the 
complex action problem \cite{AANV02}. 
In fact, at $\mu=0$ the situation is analogous: while most tests
of matrix model  
predictions have been performed only for pure gauge theories, 
dynamical fermions are also well described \cite{BMW98}. 
For a review of matrix models in QCD we refer to 
Ref.~\cite{VW00}.

\section{Matrix model predictions}

In Ref.~\cite{A02} a matrix model has been formulated in terms of $N$
complex eigenvalues
and solved in the limit $N\propto V\to\infty$. 
It has the same global symmetries as QCD at $\mu\neq0$ due to an equivalence
\cite{A03} to the matrix model of Ref.~\cite{Step96} at small 
$\mu$ in the phase with broken chiral symmetry. We give here only the 
quenched results for the microscopic density and refer to
Ref.~\cite{A02} for more details and higher-order correlation functions.

In Ref.~\cite{A02} two different large-$N$ limits
were found, those of weak and of strong nonhermiticity. 
In QCD at $\mu=0$, the lowest Dirac eigenvalues 
scale with $1/V$  to build up a finite condensate according 
to the Banks-Casher relation. In the weak-nonhermiticity limit, first 
introduced in \cite{FKS97}, this scaling remains unchanged. 
The support of the density remains quasi one-dimensional as we send $\mu\to0$,
keeping 
$\lim_{N\to\infty}2N\mu^2 \equiv \alpha^2$ fixed. In the strong-nonhermiticity
limit, the eigenvalues fill a two-dimensional surface, and thus the scaling 
is modified to $1/\sqrt{V}$. 
We find for the quenched microscopic density at weak nonhermiticity 
in the sector of topological charge $\nu$ \cite{A02}
\begin{eqnarray}
\label{rhoweak}
  \rho_{\rm weak}(\xi) &=& \frac{\sqrt{\pi\alpha^2}}
  {\mbox{erf}(\alpha)}\ |\xi|\ \exp\!\left[-\frac{(\Im
    m\,\xi)^2}{\alpha^2}\right] \\
&&\times\int_0^1 dt\ \mbox{e}^{-\alpha^2t}
  J_{|\nu|}(\sqrt{t}\xi)J_{|\nu|}(\sqrt{t}\xi^*)\:.\nonumber
\end{eqnarray}
Here, we have rescaled the complex eigenvalues according to $\xi\sim Vz$.
At strong nonhermiticity, $\mu$ is kept fixed 
as $V\to\infty$, and one obtains \cite{A02}
\begin{equation}
\label{rhostrong}
  \rho_{\rm strong}(\xi) = \sqrt{\frac{\pi}{\mu^2}}\ 
   |\xi|\ \exp\!\left[\frac{-|\xi|^2}{2\mu^2}\right] 
I_{|\nu|}\!\left(\frac{|\xi|^2}{2\mu^2}\right),
\end{equation}
rescaling the eigenvalues according to $\xi\sim\sqrt{V}z$. 
The microscopic density Eq.~(\ref{rhostrong}) is rotationally invariant in 
contrast to the macroscopic support.

\section{Lattice Data at $\mu\neq0$}

\begin{figure}[-t]
  \begin{center}
    \includegraphics[height=30.3mm]{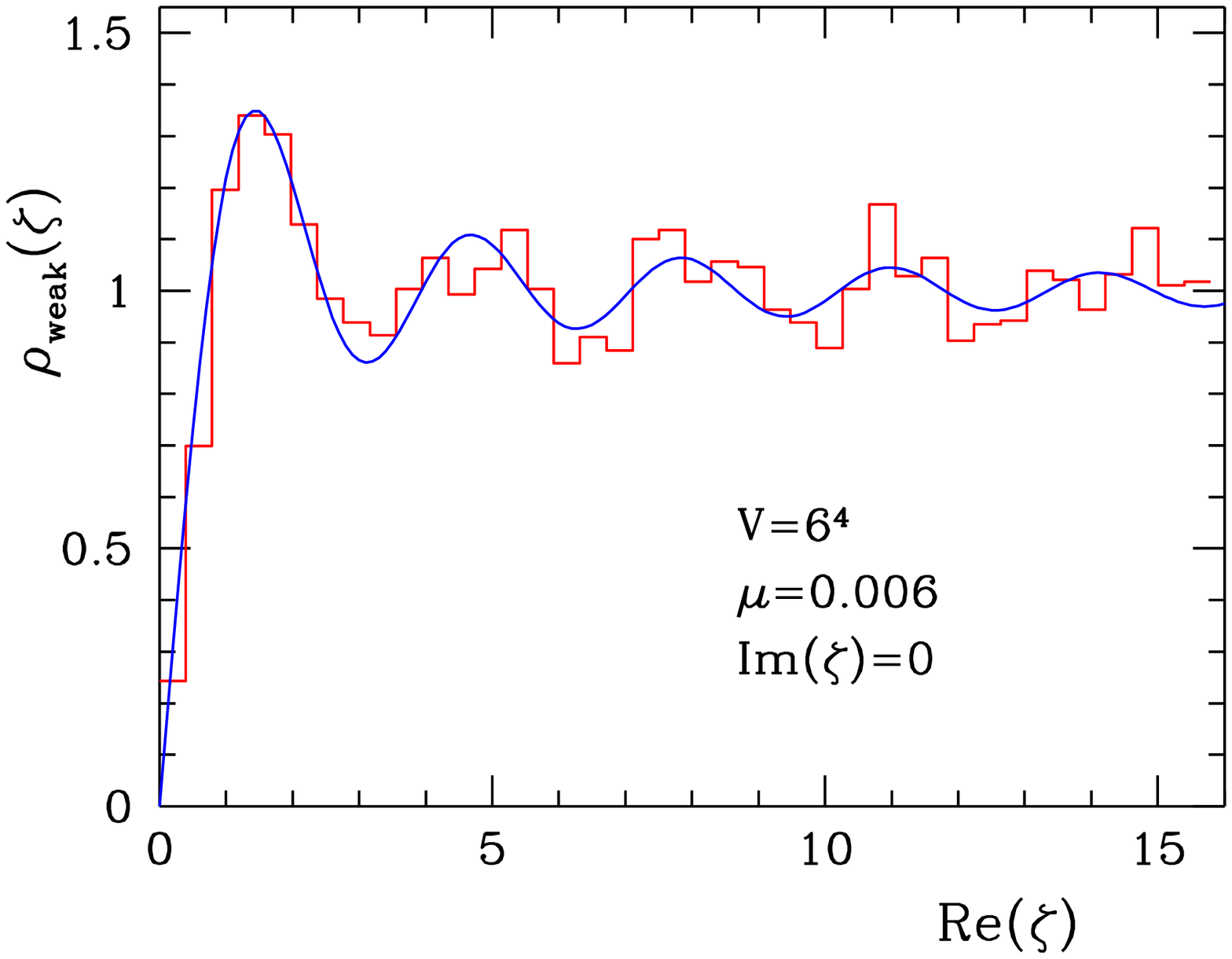}
    \includegraphics[height=30.3mm]{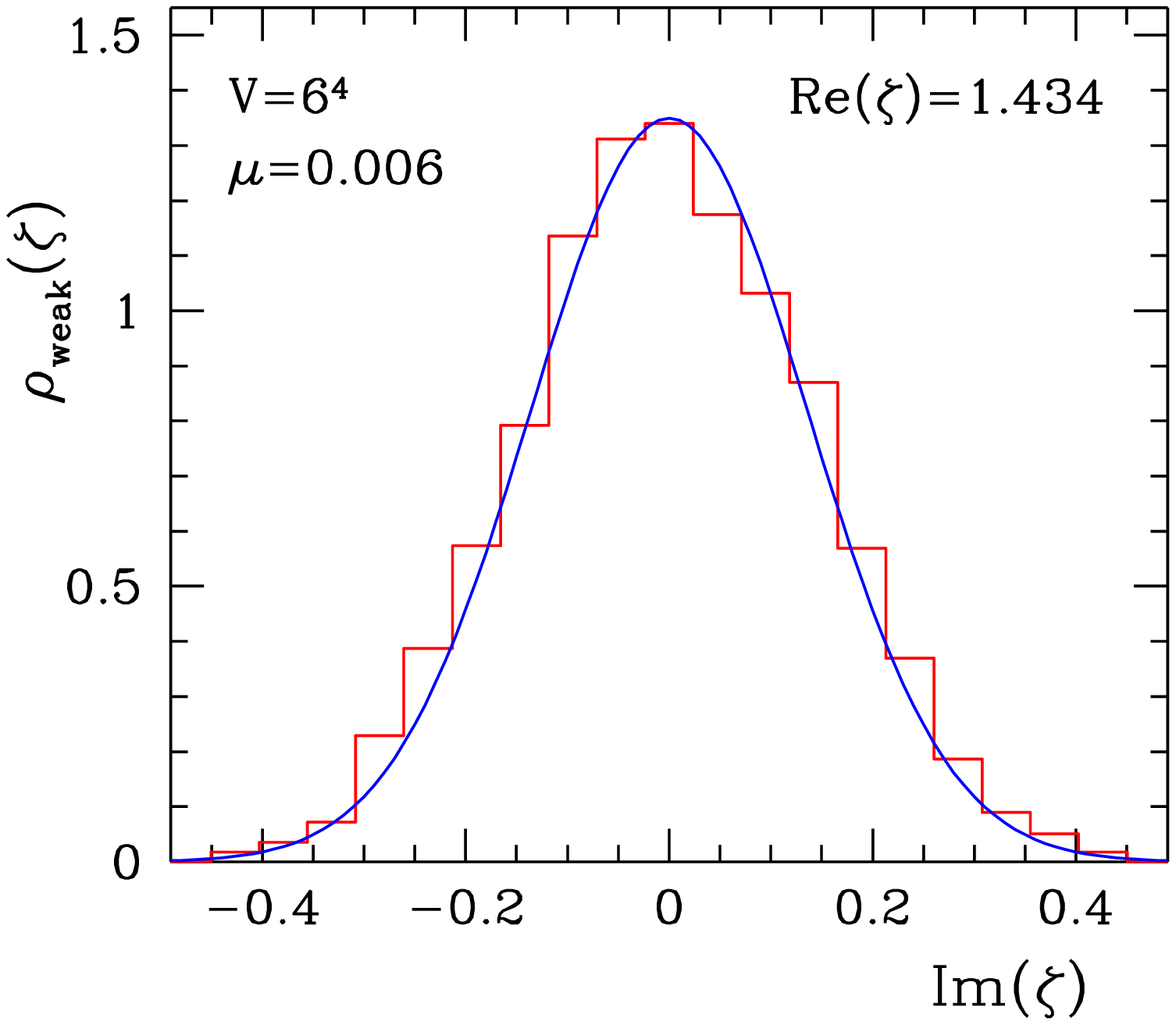}\\[3mm]
    \includegraphics[height=30.3mm]{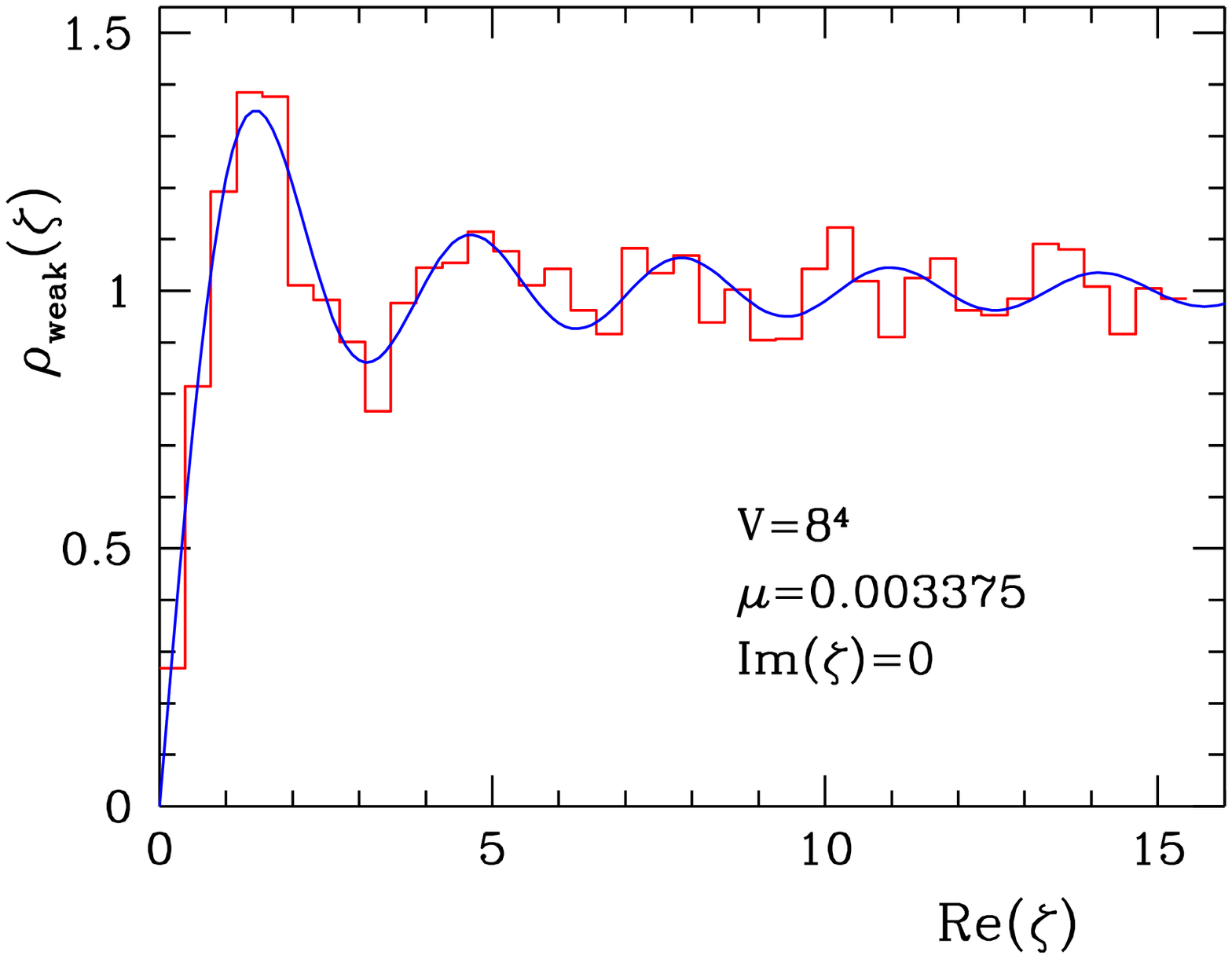}
    \includegraphics[height=30.3mm]{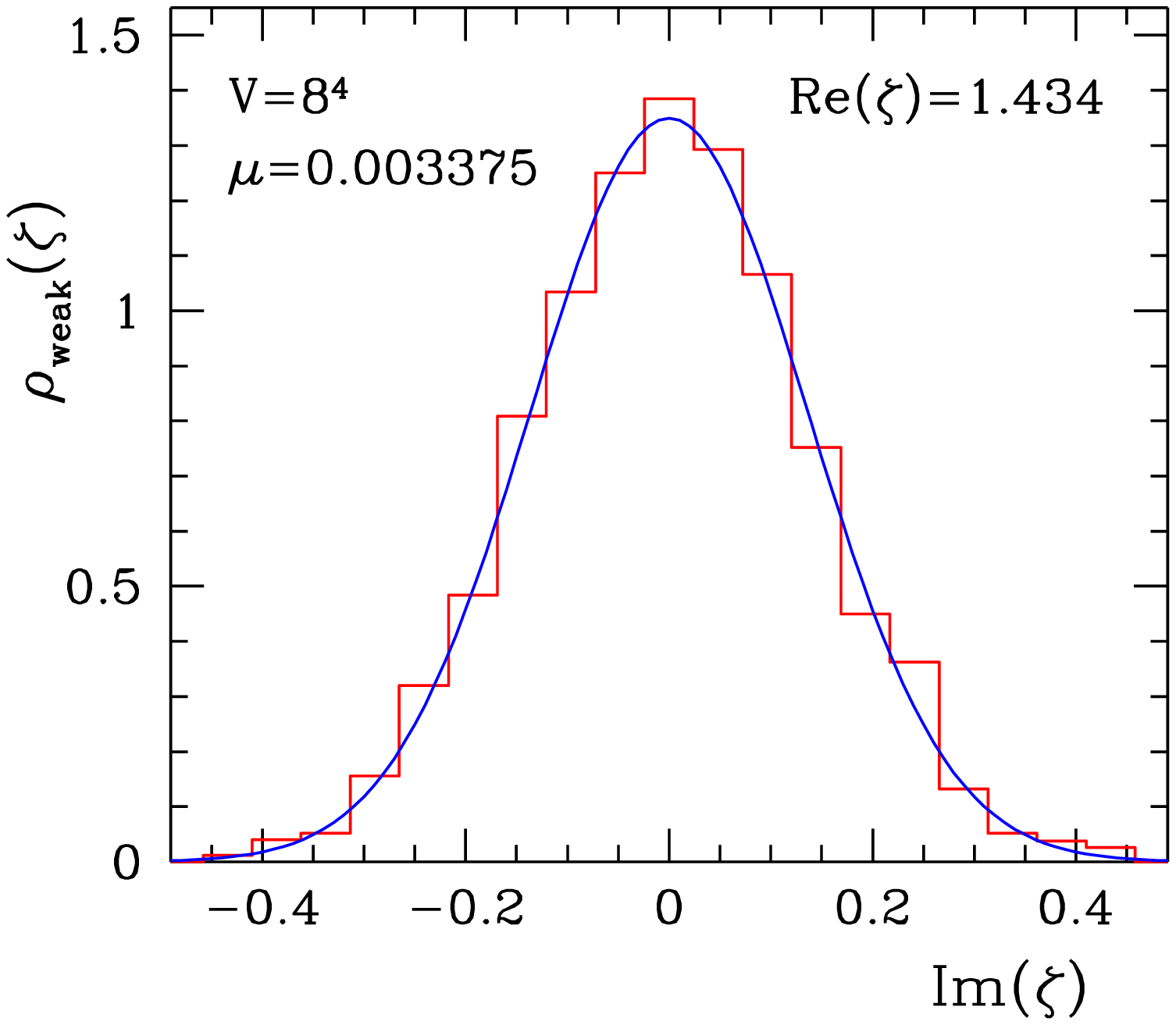}\\[3mm]
    \includegraphics[height=30.3mm]{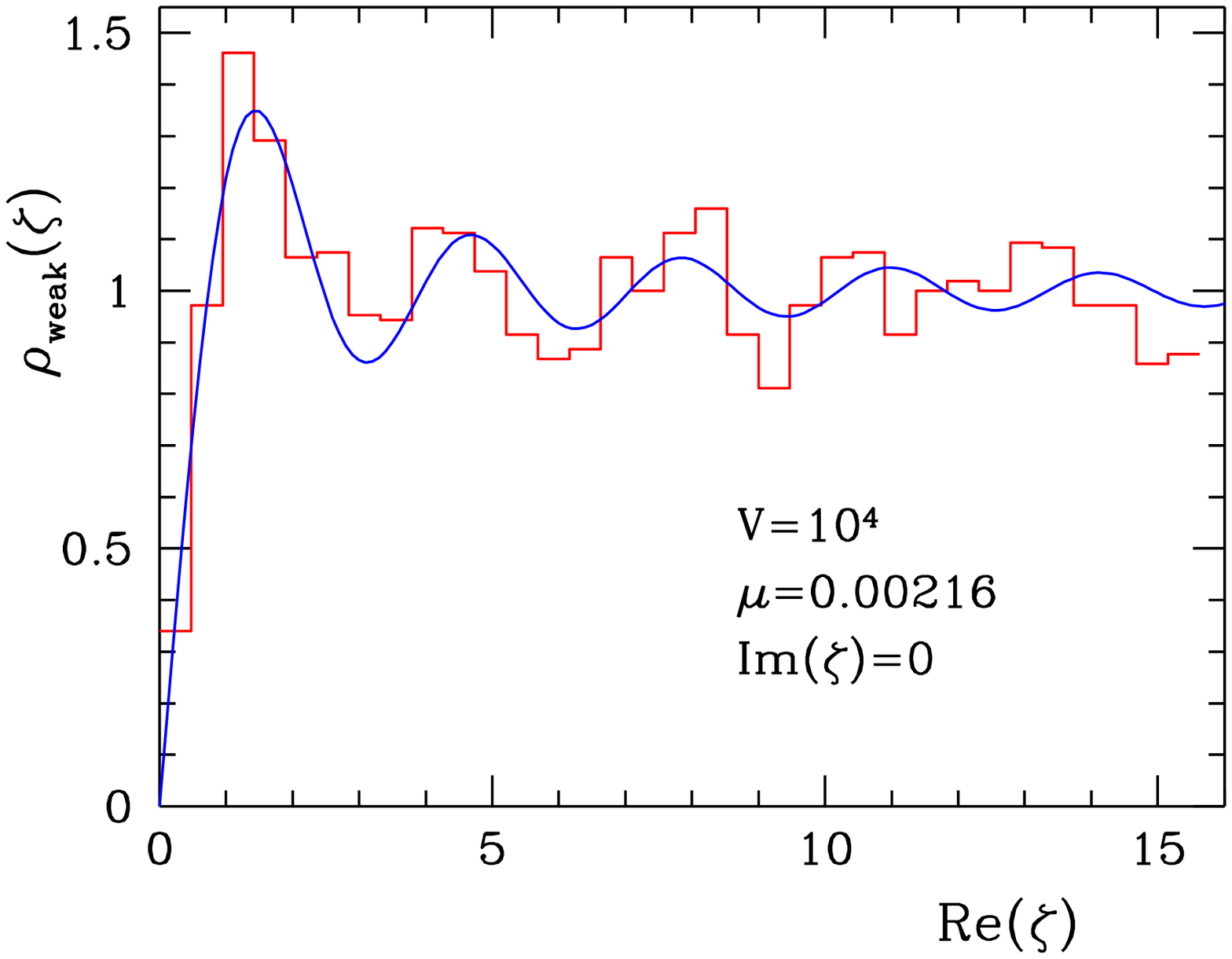}
    \includegraphics[height=30.3mm]{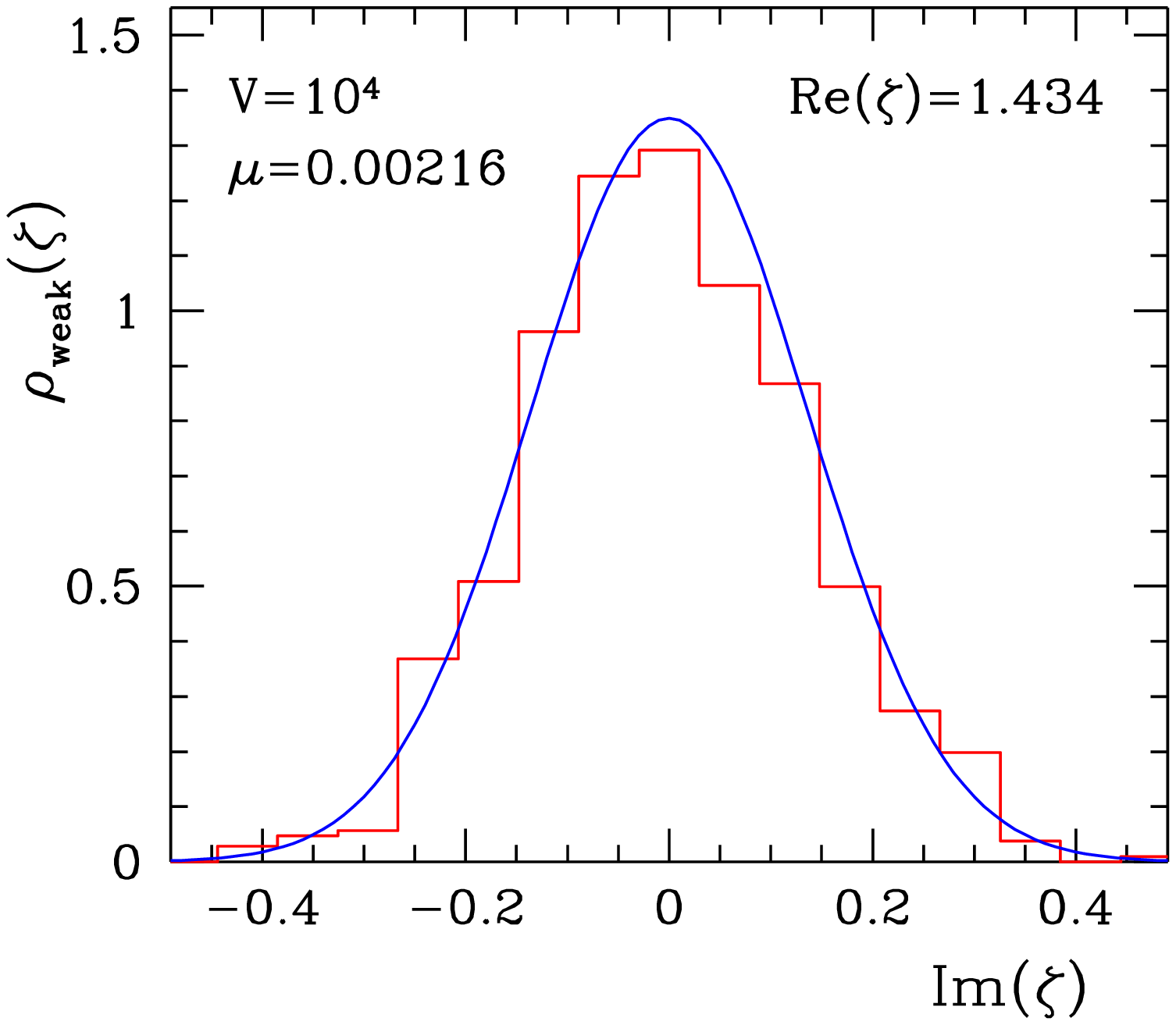}
  \end{center}
\vspace*{-8mm}
\caption{Densities of small Dirac eigenvalues 
cut along the real axis (left) and parallel to the imaginary axis at the
first maximum (right).  Data (histograms) for $V=6^4$ at $\mu=0.006$ 
(top), $V=8^4$ at $\mu=0.003375$ (middle), and $V=10^4$ at
$\mu=0.00216$ (bottom) 
vs.\ Eq.~(\ref{rhoweak}) for weak nonhermiticity.}
\label{weakdata}
\end{figure}

\begin{figure}[-t]
  \begin{center}
    \includegraphics[height=30.3mm]{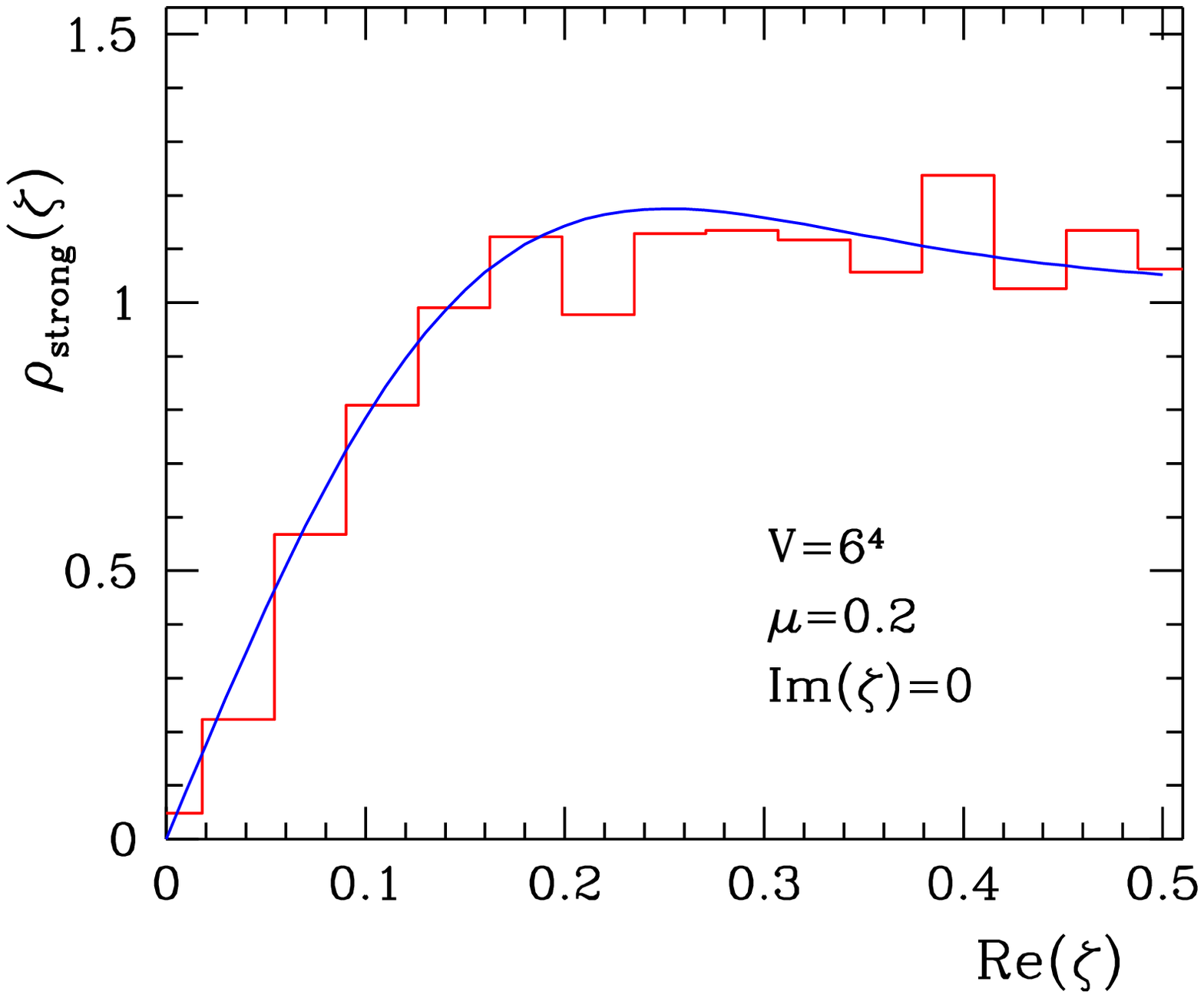}
    \includegraphics[height=30.3mm]{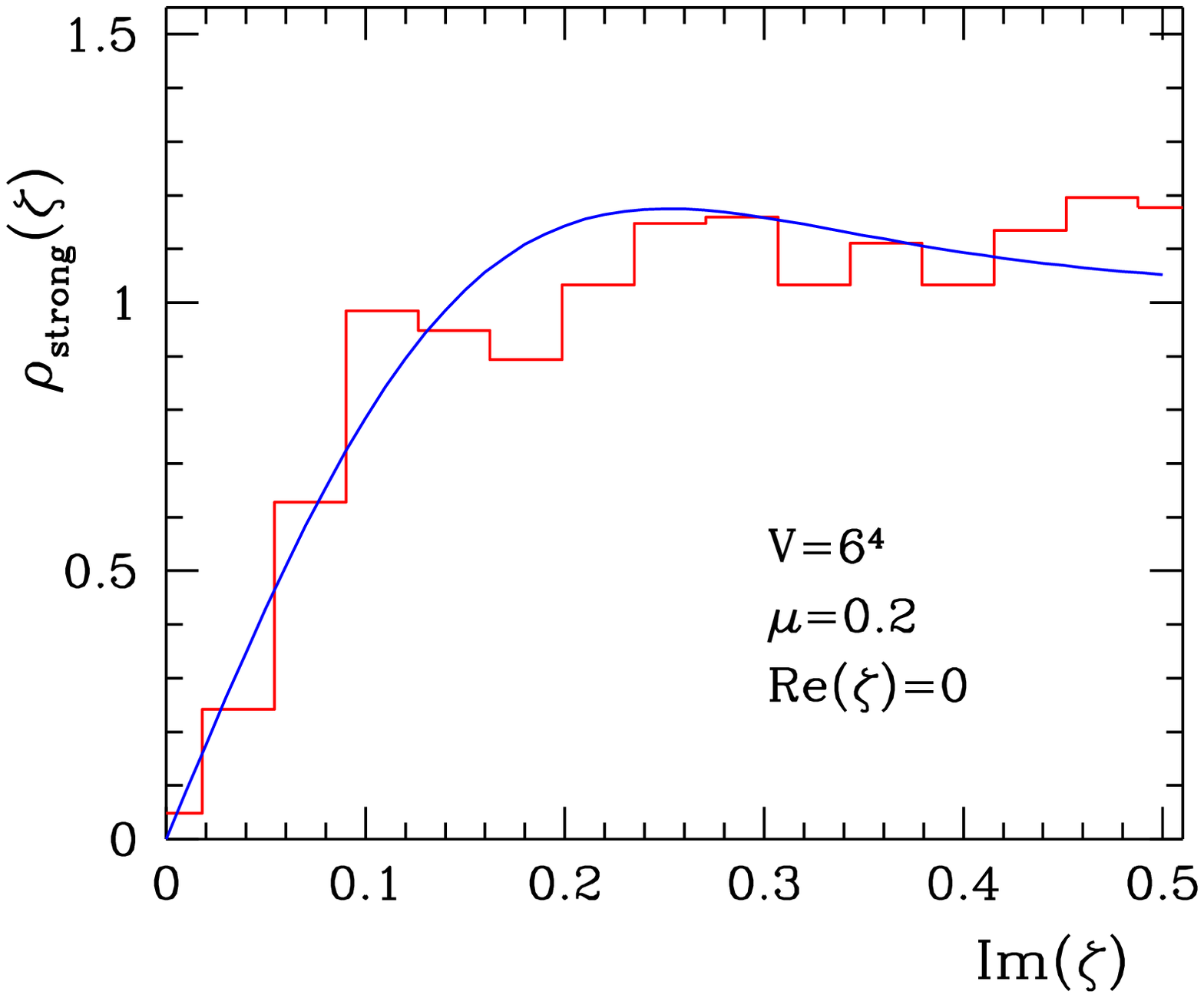}\\[3mm]
    \includegraphics[height=30.3mm]{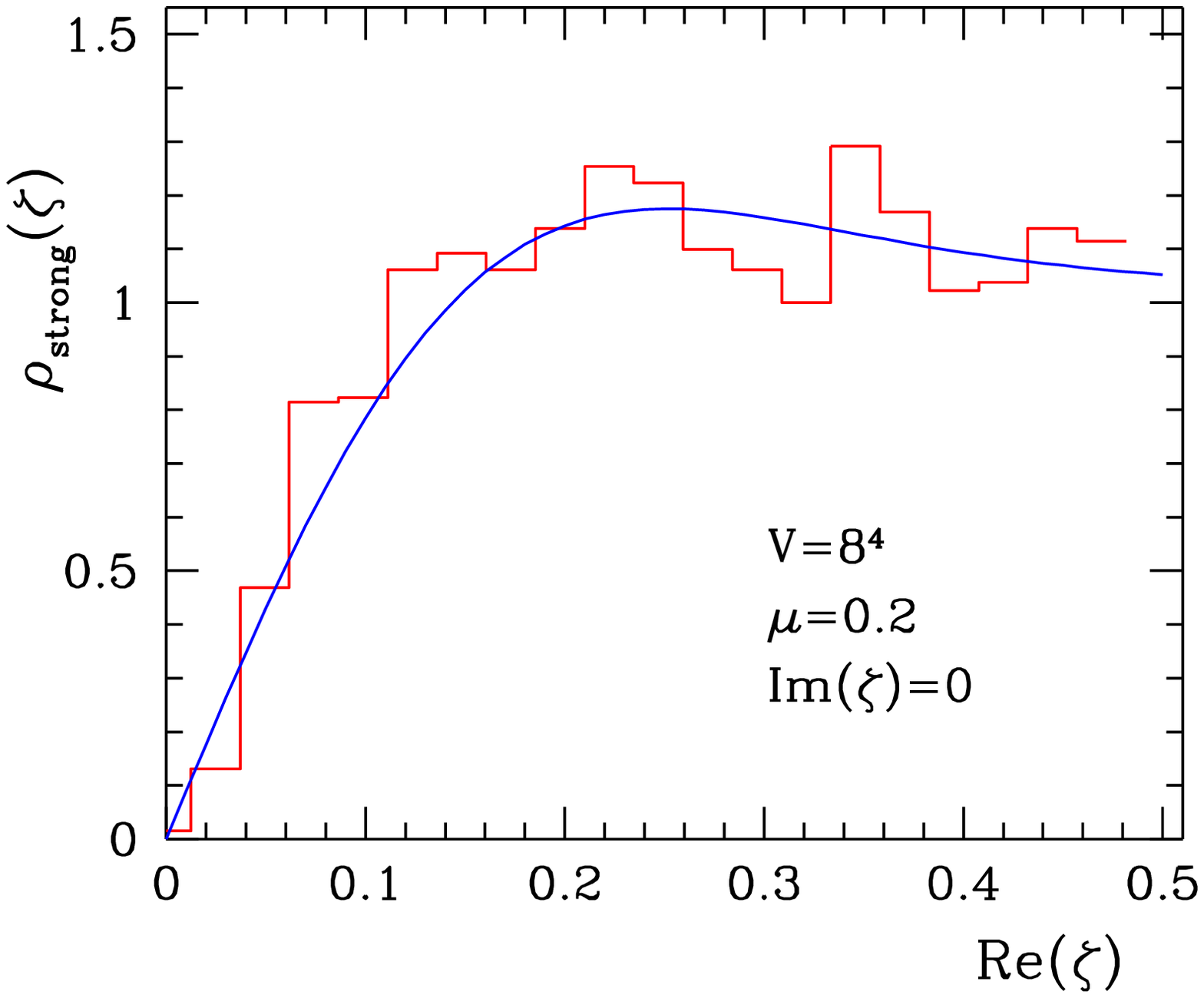}
    \includegraphics[height=30.3mm]{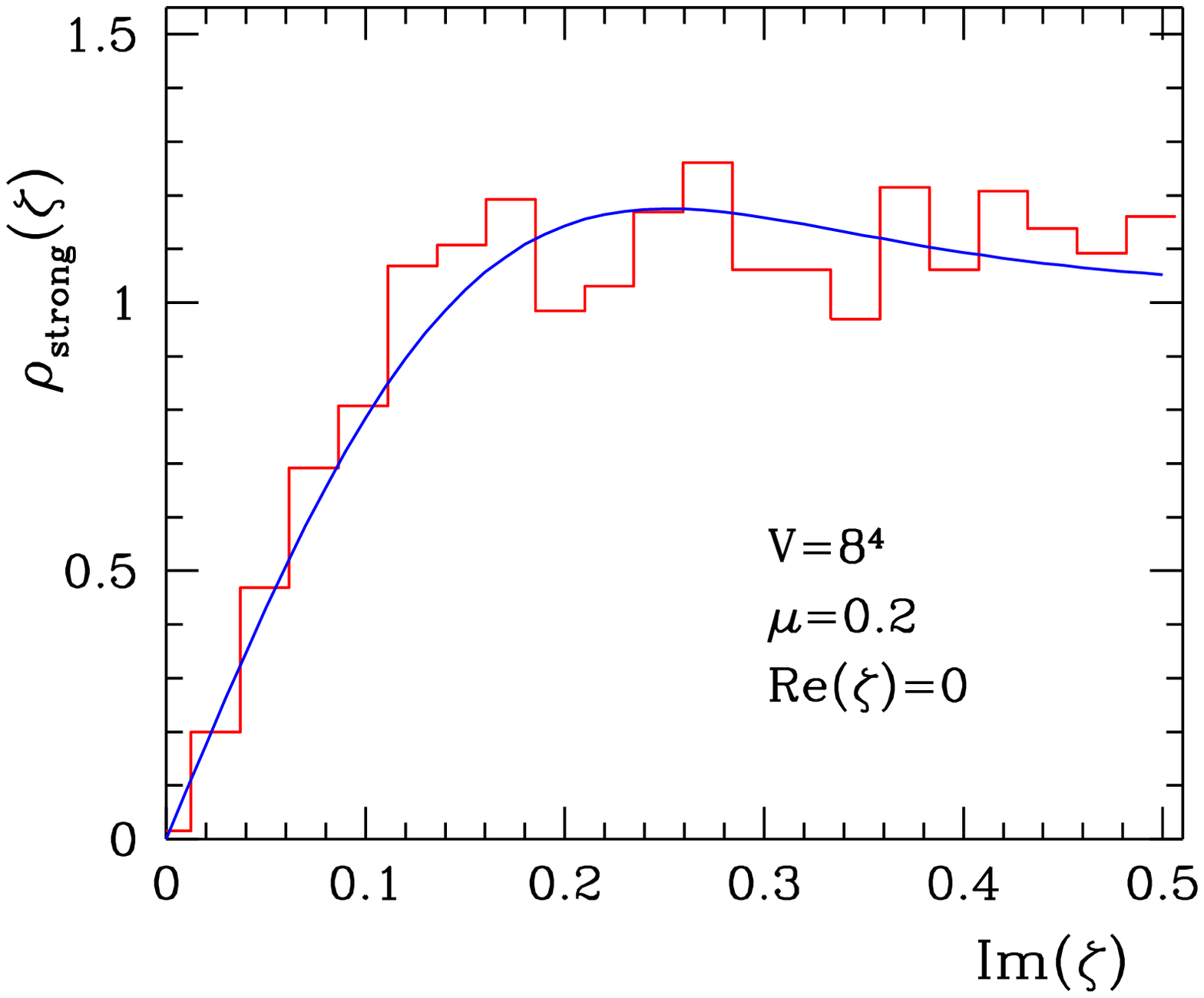}\\[3mm]
    \includegraphics[height=30.3mm]{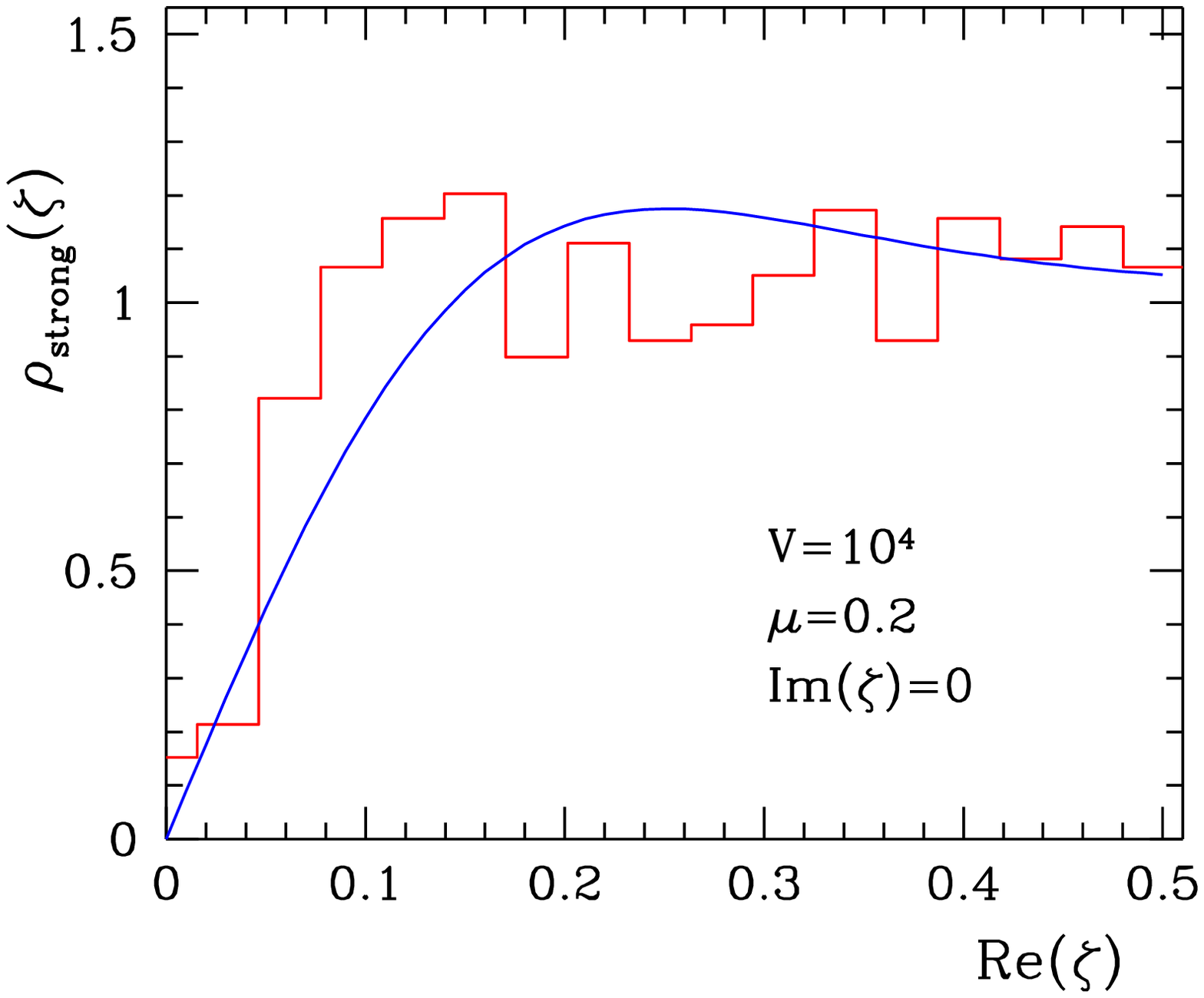}
    \includegraphics[height=30.3mm]{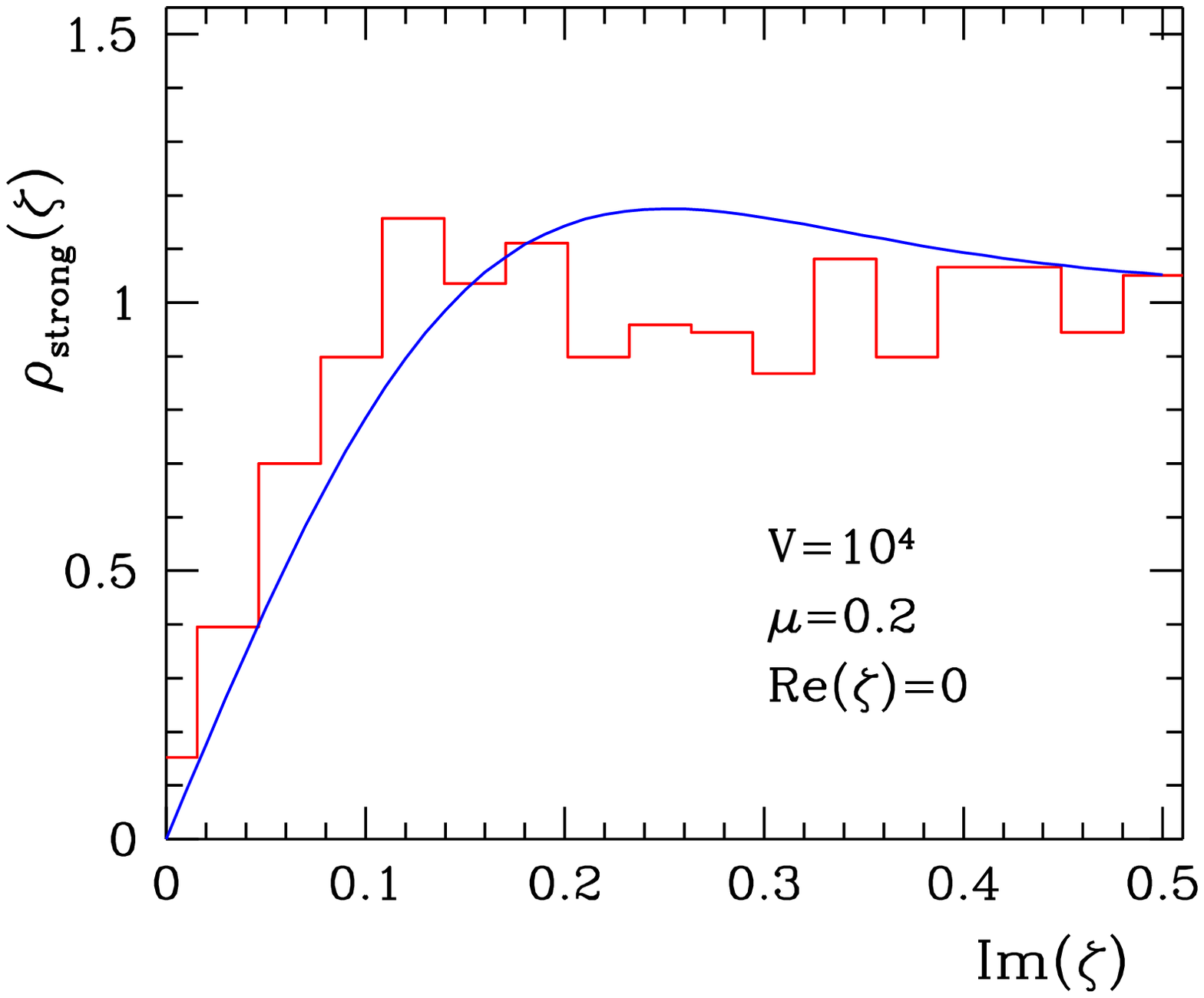}
  \end{center}
\vspace*{-8mm}
\caption{Densities of small Dirac eigenvalues 
cut along the real (left) and imaginary 
(right) axes.  Data (histograms) for $V=6^4$ (top),   $V=8^4$ (middle),
and  $V=10^4$ (bottom), all at $\mu=0.2$, 
vs. Eq.~(\ref{rhostrong}) for strong nonhermiticity.}
\label{strongdata}
\end{figure}

Our lattice calculations were done using the staggered Dirac operator 
with chemical potential $\mu\neq0$ for a pure SU(3) gauge theory, 
corresponding to $N_f=0$. There are two reasons for this choice. 
First, we need high statistics ($\approx 20,000$
configurations\footnote{The $V\!=\!10^4$ data  
in Figs.~\ref{weakdata},\ref{strongdata}
contain only 4,000 configurations, therefore the statistical errors
are larger.})
to test the matrix model predictions of Eqs.~(\ref{rhoweak},\ref{rhostrong}).
Ginsparg-Wilson type operators would be too expensive here.
Second, Wilson fermions break chiral symmetry explicitly and have
complex eigenvalues already at $\mu=0$.

In the simulations, we have chosen $\beta=6/g^2=5.0$ in the strong-coupling 
regime for the following reason. At zero temperature and $\mu=0$ the
matrix model \cite{SV93} 
is equivalent to a low-energy effective theory of QCD (the effective chiral
Lagrangian) in the so-called $\varepsilon$-regime, where the zero-momentum 
modes of the pseudo-Goldstone bosons dominate (see e.g.~\cite{VW00}). 
When the nonzero-momentum 
modes start contributing, this equivalence breaks down. The
corresponding scale, 
the so-called Thouless energy, is a function of both $\beta$ and $V$. 
As $\beta$ is increased, fewer eigenvalues are described by the matrix model, 
an effect that can be compensated by increasing $V$. 
For our small volumes $V=6^4,8^4,10^4$, however, we are limited to
relatively small values of $\beta$
(assuming a similar behavior of the Thouless energy
for small $\mu\neq0$). 

At strong coupling and thus away from the continuum limit, 
staggered fermions have the disadvantage of shifting the (topological)
Dirac zero modes 
and mixing them with the nonzero modes. We have accounted for this 
by setting $\nu=0$ in 
Eqs.~(\ref{rhoweak}) and (\ref{rhostrong}) above. 

The three different lattice volumes are chosen in order 
to test the different scaling behavior of eigenvalues and chemical potential 
for weakly and strongly nonhermitian lattice Dirac operators.
For more simulation details we refer to \cite{AW03}.

Let us first discuss the data at weak nonhermiticity. In Eq.~(\ref{rhoweak})
the average level spacing between consecutive eigenvalues is $\pi$.
In order to compare with lattice data, we first determine the mean
level spacing $d$ from the data 
averaged over many configurations and then rescale the lattice eigenvalues 
$z$ by $\xi=\pi z/d$. At the same time the spacing $d\propto 1/V$ provides us 
with the weak nonhermiticity parameter $\alpha=\mu\,\sqrt{2/d}\approx0.19$ 
to be used in 
Eq.~(\ref{rhoweak}). After normalizing the histograms to unity we thus 
obtain the parameter-free comparison shown in Fig. \ref{weakdata}. 
The three different values of $\mu$ are such that the product $V\mu^2$ 
is constant. All data are well described by the same value\footnote{
The values of $\alpha$ obtained from the $V=8^4$ and $10^4$ data agree
within errors.}
of $\alpha$ determined from the volume $V=6^4$.
Apart from the scaling $V\mu^2\propto\alpha^2$ we have thus also confirmed the 
scaling of the complex Dirac eigenvalues with the volume.

At strong nonhermiticity we have chosen $\mu=0.2$ for all lattice volumes. 
In this case, the eigenvalues are rescaled with the square root 
of the volume, $\xi=\pi z/\sqrt{d}$, where the level spacing is now 
obtained by averaging over the smallest geometric distance between 
complex eigenvalues. 
The normalized data confirm the prediction of Eq.~(\ref{rhostrong})
together with the different scaling.

In conclusion, we have confirmed the analytical matrix model predictions 
in two different scaling regimes from quenched QCD lattice data. 
For the small values of $\mu$ at weak 
nonhermiticity, unquenched simulations should be feasible
as well, keeping $V\mu^2$ fixed.
On the other hand, the lattice simulations \cite{FK02}
close to the phase transition are 
at strong nonhermiticity, since $\mu$ is kept constant.  
So far, our analytical predictions are restricted to the broken phase
and, unfortunately, are not suitable to describe the transition region.
It would be very interesting to compute predictions for the unbroken
phase as well.

\end{document}